\documentclass{moriond}

\usepackage{bm}
\usepackage{xspace}
\usepackage{amsmath}

\def\be{\begin{equation}}
\def\ee{\end{equation}}
\def\bea{\begin{eqnarray}}
\def\eea{\end{eqnarray}}

\newcommand{\comment}[1]{}

\newcommand{\mumu}{\ensuremath{\mu^+\mu^-}}
\newcommand{\ellell}{\ensuremath{\ell^+\ell^-}}
\newcommand{\epem}{\ensuremath{e^+e^-}}
\newcommand{\qsq}{\ensuremath{q^2}}
\newcommand{\Bu}{\ensuremath{B^+}}
\newcommand{\Bd}{\ensuremath{B^0}}
\newcommand{\Bs}{\ensuremath{B^0_s}}
\newcommand{\Kp}{\ensuremath{K^+}}

\newcommand{\pim}{\ensuremath{\pi^-}}
\newcommand{\Kstarz}{\ensuremath{K^{*0}}}
\newcommand{\Kstarp}{\ensuremath{K^{*+}}}
\newcommand{\jpsi}{\ensuremath{J/\psi}}
\newcommand{\gev}{\ensuremath{\mathrm{\,Ge\kern -0.1em V}}\xspace}
\newcommand{\mev}{\ensuremath{\mathrm{\,Me\kern -0.1em V}}\xspace}
\newcommand{\kev}{\ensuremath{\mathrm{\,ke\kern -0.1em V}}\xspace}
\newcommand{\ev}{\ensuremath{\mathrm{\,e\kern -0.1em V}}\xspace}
\newcommand{\gevc}{\ensuremath{{\mathrm{\,Ge\kern -0.1em V\!/}c}}\xspace}
\newcommand{\mevc}{\ensuremath{{\mathrm{\,Me\kern -0.1em V\!/}c}}\xspace}
\newcommand{\gevcc}{\ensuremath{{\mathrm{\,Ge\kern -0.1em V\!/}c^2}}\xspace}
\newcommand{\gevgevcccc}{\ensuremath{{\mathrm{\,Ge\kern -0.1em V^2\!/}c^4}}\xspace}
\newcommand{\decay}[2]{\ensuremath{#1\!\to #2}\xspace}
\newcommand{\deriv}{\ensuremath{\mathrm{d}}}
\newcommand{\invfb}{\ensuremath{\mbox{\,fb}^{-1}}\xspace}
\newcommand{\BF}{\ensuremath{{\cal B}}}
\newcommand{\stat}{\ensuremath{\mathrm{(stat)}}}
\newcommand{\syst}{\ensuremath{\mathrm{(syst)}}}
\newcommand{\RK}{\ensuremath{R_K}}
\newcommand{\RKst}{\ensuremath{R_{K^*}}}

\newcommand{\Photo}{\includegraphics[height=35mm]{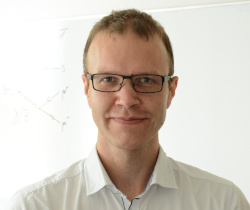}}

\begin{document}
\vspace*{4cm}
\title{{{\bm{$b\to s\ell\ell$}}} and Lepton Universality at LHCb}

\author{C.~Langenbruch on behalf of the LHCb collaboration}

\address{
  Physikalisches Institut,
  Heidelberg University, INF 226, 69120 Heidelberg, Germany
}

\maketitle\abstracts{
Lepton universality, meaning the equal coupling of the electroweak gauge bosons to the different lepton flavours, is a central property of the Standard Model (SM). 
Tests of lepton universality in ratios of rates of $b\to s\ell^+\ell^-$ decays
involving electron and muon final states 
profit from precise SM predictions, 
free from hadronic uncertainties. 
They therefore constitute particularly powerful probes for New Physics (NP) scenarios. 
These proceedings summarise the latest and most precise tests of lepton universality in rare $b\to s\ell^+\ell^-$ transitions by the LHCb collaboration.
}

\section{Introduction}
\begin{figure}
\begin{minipage}{0.495\linewidth}
\centerline{\includegraphics[width=0.7\linewidth]{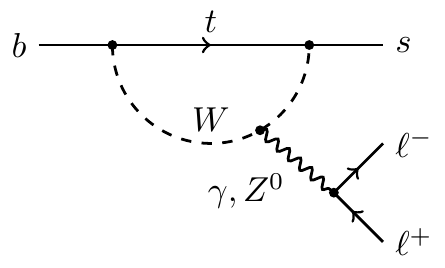}}
\end{minipage}
\hfill
\begin{minipage}{0.495\linewidth}
\centerline{\includegraphics[width=0.7\linewidth]{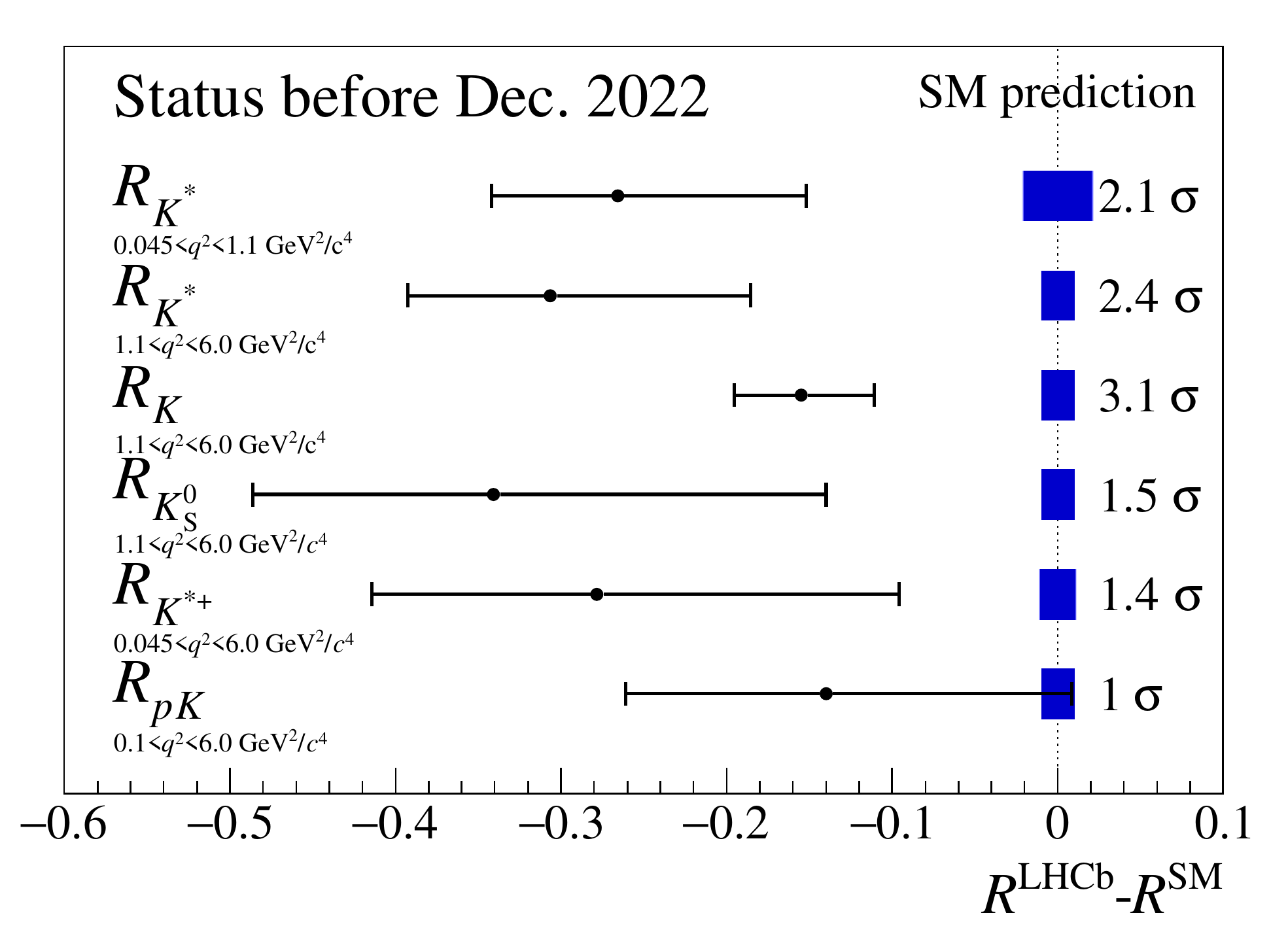}}
\end{minipage}
\caption[]{(Left) Lowest order $b\to s\ell^+\ell^-$ transition in the SM. (Right) Experimental status before December 2022, showing measurements of (from top to bottom) $R_{K^*}$~\cite{LHCb:2017avl}, $R_K$~\cite{LHCb:2021trn}, $R_{K_S^0}$~\cite{LHCb:2021lvy}, $R_{K^{*+}}$~\cite{LHCb:2021lvy} and $R_{pK}$~\cite{LHCb:2019efc} by the LHCb collaboration. The reported tensions with the SM predictions are also given in terms of standard deviations.} 
\label{fig:introduction}
\end{figure}

Rare $\decay{b}{s\ellell}$ decays constitute sensitive probes for NP,
as they are forbidden at tree-level in the SM and only allowed at loop order, see Fig.~\ref{fig:introduction} (left). 
Rare decays are therefore both loop- and CKM suppressed and virtual NP contributions can have a large impact, 
resulting in significant deviations of physical observables from their SM predictions. 

Recent measurements by the LHCb collaboration have shown some interesting tensions with SM predictions,
among them in measurements of branching fractions and angular observables of $b\to s\mumu$ processes.
The branching fraction measurements, most notably of the decays $\decay{\Bu}{\Kp\mumu}$, $\decay{\Bd}{\Kstarz\mumu}$ and $\decay{\Bs}{\phi\mumu}$, consistently lie below the SM predictions~\cite{LHCb:2014cxe,LHCb:2016ykl,LHCb:2021zwz,LHCb:2015tgy}, with tensions corresponding to $1\textrm{--}3$ standard deviations ($\sigma$). 
The tensions in angular analyses are particularly pronounced in the angular observable $P_5^\prime$, 
the measurements in the decays $\decay{\Bd}{\Kstarz\mumu}$ and $\decay{\Bu}{\Kstarp\mumu}$ are found to be in tension with the SM prediction at around $3\,\sigma$ each~\cite{LHCb:2020lmf,LHCb:2020gog}. 
The decay $\decay{\Bs}{\phi\mumu}$ does not allow access to $P_5^\prime$ but shows consistent, but less significant, tensions~\cite{LHCb:2021xxq}. 
However, the SM predictions of both branching fractions and angular observables are affected by hadronic uncertainties;
the angular observables by $c\bar{c}$-loop contributions, which are difficult to estimate in theory,
and the branching fractions in addition by form-factors, which require non-perturbative calculations. 
The exact significance of the tensions observed in $\decay{b}{s\mumu}$ decays is therefore currently under discussion. 

Tests of lepton universality allow to 
search for NP effects without being affected by potential hadronic uncertainties of the SM prediction. 
Lepton universality ratios in $\decay{b}{s\ellell}$ decays defined as 
\begin{equation}
  R_X(\qsq_{\rm min},\qsq_{\rm max}) = \frac{\int_{\qsq_{\rm min}}^{\qsq_{\rm max}}\frac{\deriv\Gamma(\decay{B}{X_s\mumu})}{\deriv\qsq}\deriv\qsq}{\int_{\qsq_{\rm min}}^{\qsq_{\rm max}}\frac{\deriv\Gamma(\decay{B}{X_s\epem})}{\deriv\qsq}\deriv\qsq},\label{eq:rkdefinition}
\end{equation}
are predicted to be exactly unity in the SM in \qsq\ regions where the lepton masses can be neglected. 
Here, \qsq\ is defined as the invariant mass of the dilepton system squared. 
The leading uncertainty of the SM prediction arises from QED effects and is at most of the order of $1\,\%$~\cite{Bordone:2016gaq}. 
Potential hadronic contributions would affect the electron and muon mode the same way, and thus cancel in the ratio. 

The experimental status before December 2022 is shown in Fig.~\ref{fig:introduction} (right). 
The LHCb collaboration performed measurements of
$R_{K^*}$~\cite{LHCb:2017avl}, $R_K$~\cite{LHCb:2021trn}, $R_{K_S^0}$~\cite{LHCb:2021lvy}, $R_{K^{*+}}$~\cite{LHCb:2021lvy} and $R_{pK}$~\cite{LHCb:2019efc}, 
which seemed to consistently lie below the SM prediction, with quoted significances of up to $3.1\,\sigma$~\cite{LHCb:2021trn}. 
These proceedings focus on the most recent test of lepton universality performed by the LHCb collaboration which appeared in December 2022~\cite{LHCb:2022qnv,LHCb:2022zom}, 
a simultaneous determination of $R_K$ and $R_{K^*}$ using the full data set taken by LHCb during the LHC Run~1 and 2, corresponding to an integrated luminosity of $9\invfb$. 

\section[Simultaneous determination of $R_K$ and $R_{K^*}$]{Simultaneous determination of {\bm{$R_K$}} and {\bm{$R_{K^*}$}}}
\begin{figure}
\centerline{\includegraphics[width=0.7\linewidth]{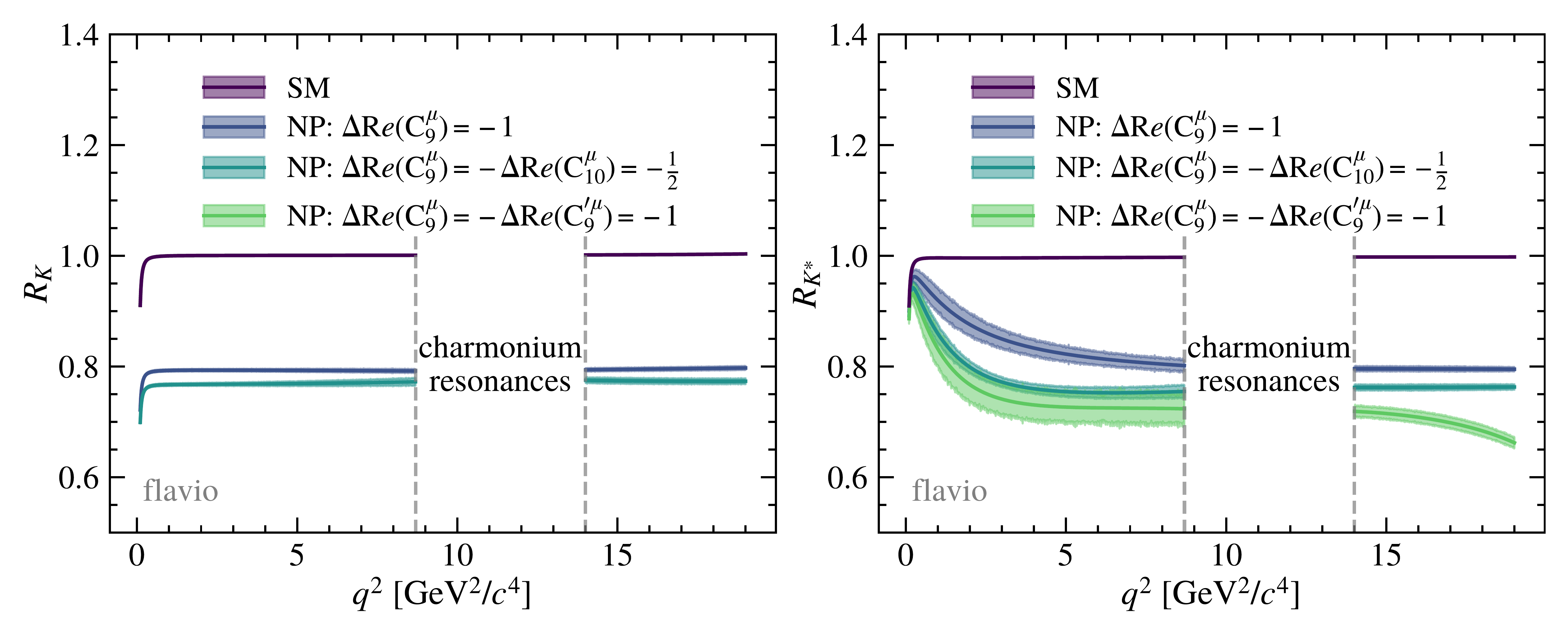}}
\caption[]{(Left) $R_K$ and (right) $R_{K^*}$ in the SM and for several NP scenarios, assuming different deviations of the muonic effective vector (axial-vector) coupling $C_9$ ($C_{10}$) from the SM values. Figure taken from Ref.~\cite{LHCb:2022qnv,LHCb:2022zom}.} 
\label{fig:npscenarios}
\end{figure}

A simultaneous determination of $R_K$ and $R_{K^*}$ allows
to search for potential NP effects in decays of $B$ mesons to pseudoscalar- and to vector-mesons,
thereby providing some ability to disentangle different NP models.
This is illustrated in~Fig.~\ref{fig:npscenarios}, which shows predictions for $R_K$ and $R_{K^*}$ in the SM and for several NP scenarios. 
In addition, Fig.~\ref{fig:npscenarios} shows that it is beneficial to measure the lepton flavour universality ratios in several \qsq\ regions, 
this analysis measures $R_K$ and $R_{K^*}$ in two \qsq\ ranges,
in the low-\qsq\ region $0.1<\qsq<1.0\gevgevcccc$ and the central-\qsq\ region $1.1<\qsq<6.0\gevgevcccc$. 

\subsection{Analysis strategy}
Like previous measurements of lepton flavour universality, 
the analysis leverages double ratios, expanding the definition of $R_K$ in Eq.~\ref{eq:rkdefinition} by the branching fraction ratio of the tree-level control and normalisation mode $\decay{\Bu}{\jpsi\Kp}$, resulting in 
\begin{eqnarray}
  R_{K} &=& \frac{\BF(\decay{\Bu}{\Kp\mumu})}{\BF(\decay{\Bu}{\Kp\epem})} \times \frac{\BF(\decay{\Bu}{\Kp\jpsi_{\to \epem}})}{\BF(\decay{\Bu}{\Kp\jpsi_{\to\mumu}})} \label{eq:doubleratio}\\
  &=& 
  \frac{N_{B\to K\mu\mu}}{N_{B\to Kee}}\times \frac{N_{B\to K\jpsi_{\to ee}}}{N_{B\to K\jpsi_{\to \mu\mu}}} \times
  \frac{\epsilon_{B\to Kee}}{\epsilon_{B\to K\jpsi_{\to ee}}}\times \frac{\epsilon_{B\to K\jpsi_{\to \mu\mu}}}{\epsilon_{B\to K\mu\mu}}, 
  \nonumber
\end{eqnarray}
and an analogous expression for the $R_{K^*}$ ratio. 
The efficiencies $\epsilon$ are taken from simulation, corrected with data-driven methods. 
Using the double ratio is experimentally advantageous, as systematic effects in the efficiencies cancel to first order in the double ratio, 
as the efficiency for the rare signal electron (muon) mode is divided by the efficiency for the electron (muon) control mode.  
This cancellation is particularly effective as the rare signal and the tree-level control mode have very similar kinematics and topology at the LHC,
since the $B$ mesons are produced strongly boosted in the lab frame. 
The double ratio approach is possible because lepton flavour universality in $\decay{\jpsi}{\ellell}$ decays is well established at the per-mille level~\cite{BESIII:2013csc}. 

\subsection{Reconstruction and selection}
The reconstruction and selection is more challenging for the rare electron modes than for the muon modes at LHCb.
This is due to lower trigger efficiencies for electrons, deteriorated momentum and subsequently mass resolution due to Bremsstrahlung, and more challenging treatment of backgrounds. 
At the trigger level, the analysis increases selection efficiency by combining candidates triggered by the signal electrons using the calorimetry 
with signal candidates that were triggered by other, non-signal, particles in the event. 
This recovers a significant number of signal events for which the electrons did not pass the trigger thresholds in the electromagnetic calorimeter. 

\begin{figure}
\centerline{\includegraphics[clip=true,trim=0mm 175mm 0mm 0mm,width=0.70\linewidth]{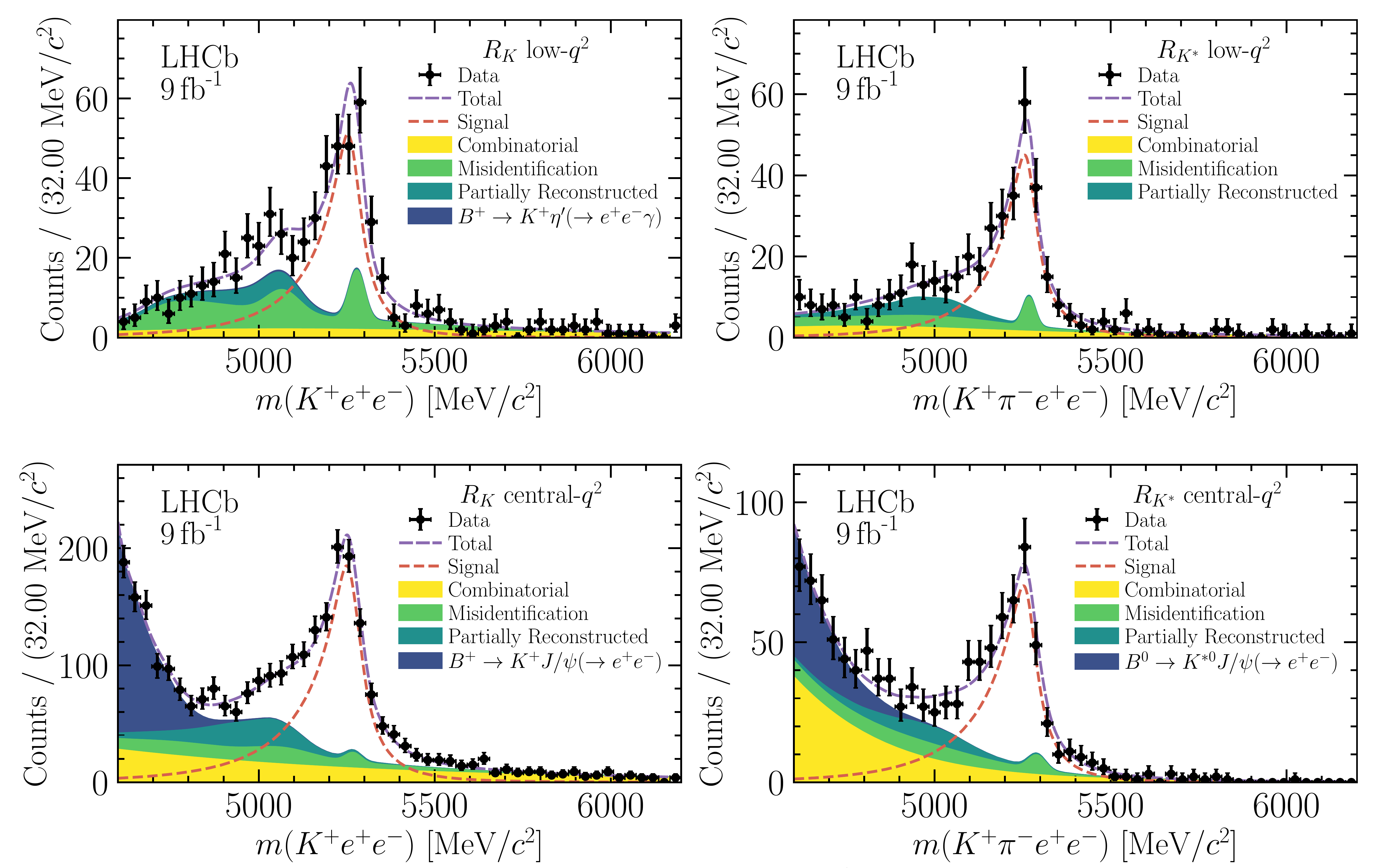}}
\caption[]{Mass distributions for (left) $K^+e^+e^-$ and (right) $K^+\pi^-e^+e^-$ final states. The figures show (top) the low \qsq\ region and (bottom) the central \qsq\ region. Figures taken from Refs.~\cite{LHCb:2022qnv,LHCb:2022zom}.} 
\label{fig:electrons}
\end{figure}

\begin{figure}
\centerline{\includegraphics[clip=true,trim=0mm 175mm 0mm 0mm,width=0.70\linewidth]{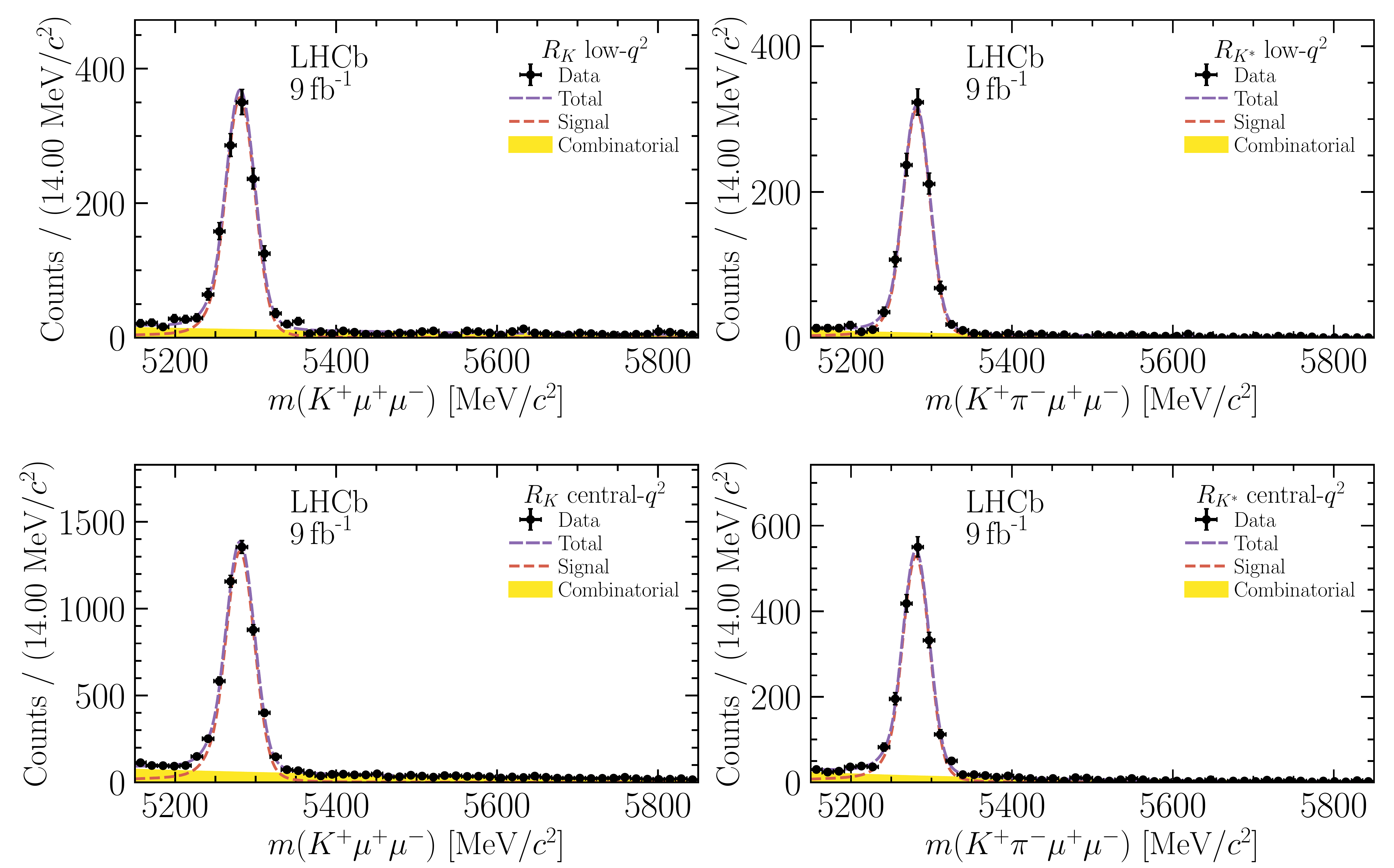}}
\caption[]{Mass distributions for (left) $K^+\mu^+\mu^-$ and (right) $K^+\pi^-\mu^+\mu^-$ final states. The figures show the (top) the low \qsq\ region and (bottom) the central \qsq\ region. Figures taken from Refs.~\cite{LHCb:2022qnv,LHCb:2022zom}.} 
\label{fig:muons}
\end{figure}

When electrons traverse material they can emit Bremsstrahlung which results in their reconstructed momentum being reduced.
The Bremsstrahlung reconstruction algorithm aims to recover this lost momentum by reconstructing photons in the electromagnetic calorimeter that are consistent with being Bremsstrahlung photons from signal electrons. 
However, even after this Bremsstrahlung reconstruction the mass resolution for the electron decays is significantly deteriorated compared to the muon modes,
which is apparent when comparing Figs.~\ref{fig:electrons} and~\ref{fig:muons}. 

The reduced mass resolution in the electron mode requires a careful treatment of pollution by potential backgrounds. 
Combinatorial backgrounds are suppressed by a multivariate classifier using kinematic and topological information.
Partially reconstructed backgrounds 
are suppressed by a multivariate classifier using track and vertex isolation variables,
and, in addition, by a requirement on the corrected mass, which exploits the fact that the flight direction reconstructed from the final state particle momenta should align with the direction from the primary production to the secondary decay vertex for signal. 
Peaking backgrounds from misidentified decays are vetoed combining particle identification criteria and kinematic variables. 

After the selection criteria, a residual background component 
from decays where one or more hadrons are misidentified as electrons, with or without additional missing energy, is still present. 
This component is explicitly modeled in the analysis using a data-driven approach; the particle identification criteria on the leptons are inverted, creating a sample enriched with backgrounds from misidentification. High statistics control samples from data are then used to weight these events according to their misidentification probability. 
This component from residual misidentified decays is described by an empirical model in the mass fit and shown as the light green shaded area in Fig.~\ref{fig:electrons}.  

\subsection{Crosschecks}
\begin{figure}
  \centerline{\includegraphics[width=0.70\linewidth]{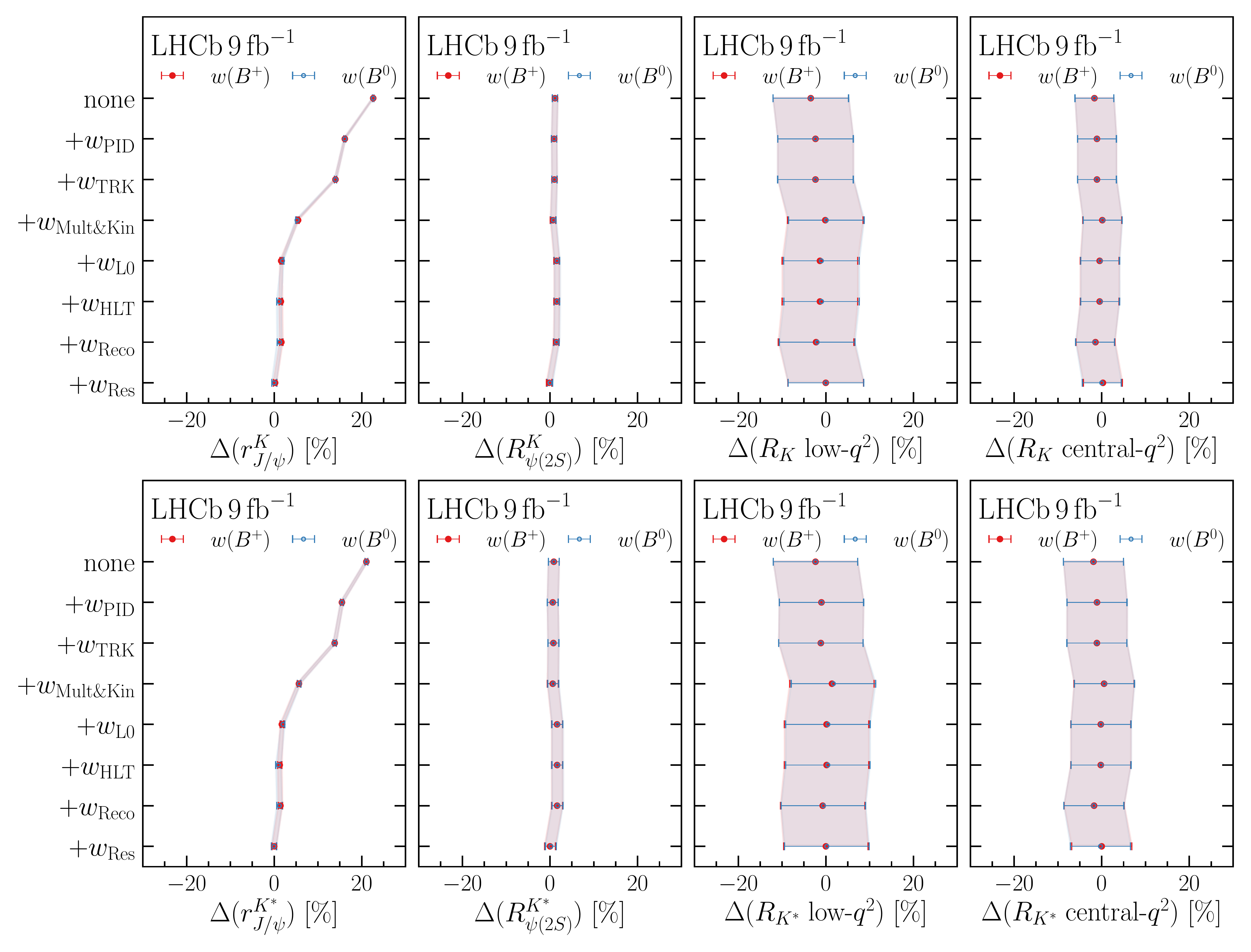}}
  \caption[]{The cross-checks $r_{\jpsi}$ and $R_{\psi(2S)}$ for the different correction steps of simulation. Deriving the simulation corrections using $\Bu$ or $\Bd$ decays gives equivalent results. Figures taken from Refs.~\cite{LHCb:2022qnv,LHCb:2022zom}.} 
\label{fig:rjpsi}
\end{figure}

The ratio of the branching fractions of the control decays, defined as
\begin{eqnarray}
  r_{\jpsi} &=& \frac{\BF(\decay{\Bu}{\Kp\jpsi_{\to \mumu}})}{\BF(\decay{\Bu}{\Kp\jpsi_{\to \epem}})} = \frac{N_{\decay{B}{K\jpsi_{\to\mu\mu}}}}{N_{\decay{B}{K\jpsi_{\to ee}}}}\times\frac{\epsilon_{\decay{B}{K\jpsi_{\to ee}}}}{\epsilon_{\decay{B}{K\jpsi_{\to \mu\mu}}}} \overset{!}{=} 1\label{eq:rjpsi}
\end{eqnarray}
for \RK, and analogously for \RKst, constitutes a stringent cross-check for the efficiencies, 
as systematic effects do not cancel in the single efficiency ratio, in contrast to the double ratio in Eq.~\ref{eq:doubleratio}. 
The integrated single ratio is found to be compatible with unity.
In addition, $r_{\jpsi}$ is found to be stable as a function of several kinematic and geometric variables. 
The simulation corrections, which include
data-driven corrections to particle identification, tracking and trigger efficiency, and reconstruction effects, 
are thus successfully validated.
A cross-check of the double-ratio approach at a different \qsq-value is provided by the quantity $R_{\psi(2S)}$,
which is defined by replacing the rare mode $\decay{\Bu}{\Kp\ellell}$ in Eq.~\ref{eq:doubleratio} with the tree-level decay $\decay{\Bu}{\Kp\psi(2S)_{\to \ellell}}$ (analogously for \RKst). 
The expectation for the double ratio $R_{\psi(2S)}$ is also unity, which is confirmed on data. 
Both crosschecks $r_{\jpsi}$ and $R_{\psi(2S)}$ are shown in Fig.~\ref{fig:rjpsi} for the different correction steps of the simulation. 
The figure illustrates the power of the double ratio, as $R_{\psi(2S)}$ is already unity without simulation corrections, whereas the single ratio $r_{\jpsi}$ requires the corrections to agree well with the expectation. 

To check the modeling of the residual misidentified backgrounds, the analysis is repeated without including them explicitly in the fit, 
but instead with progressively tightening the electron identification criteria, thereby reducing the residual pollution. 
Both $R_K$ and $R_{K^*}$ 
plateau for very tight particle identification criteria at values fully consistent with the nominal fit. 
This both illustrates the necessity of including 
residual misidentified backgrounds in the fit, 
and gives further confidence in the nominal fit results. 

\subsection{Results and discussion}
The lepton flavour universality observables \RK\ and \RKst\ at low and central \qsq\ are determined in a simultaneous fit,
using the mass distributions of the rare decays together with the corresponding normalisation modes. 
The simultaneous fit is advantageous as it allows to constrain backgrounds,  
such as the radiative tails of the \jpsi\ modes that enter the low mass region of the rare decays at central \qsq. 
Similarly, $\decay{\Bd}{\Kstarz(\to\Kp\pim)\epem}$ decays constitute a source of partially reconstructed background to the $\Kp\epem$ final state, which is constrained through the simultaneous fit. 

Systematic uncertainties can be divided into those associated with the signal and normalisation mode efficiencies,
and those associated with the the simultaneous fit. 
The dominant source of systematic uncertainty arises from the modeling of residual backgrounds from misidentification, 
overall the measurement is however strongly statistically dominated. 
The measured values for \RK\ and \RKst\ in the two studied \qsq\ ranges are 
\begin{eqnarray*}
        \RK(0.1<\qsq<1.0\gevgevcccc)   &=& 0.994~^{+0.090}_{-0.082}\,\stat 
                    \; ^{+0.029}_{-0.027}\,\syst,  \\
        \RKst(0.1<\qsq<1.0\gevgevcccc) &=& 0.927~^{+0.093}_{-0.087}\,\stat \;
                       ^{+0.036}_{-0.035}\,\syst,\\
        \RK(1.1<\qsq<6.0\gevgevcccc)  &=& 0.949~^{+0.042}_{-0.041}\,\stat\;
                      ^{+0.022}_{-0.022}\,\syst, \\
        \RKst(1.1<\qsq<6.0\gevgevcccc) &=& 1.027~^{+0.072}_{-0.068}\,\stat \;
                      ^{+0.027}_{-0.026}\,\syst,
\end{eqnarray*}
where statistical and systematic uncertainties are given separately. 
These constitute the most precise such measurements to date and are in excellent agreement with the SM. 

These results differ from previous measurements of \RK~\cite{LHCb:2021trn} and \RKst~\cite{LHCb:2017avl}. 
For \RK\ the difference is primarily due to systematic effects, specifically the treatment of backgrounds from misidentification.
The more stringent particle identification criteria 
significantly reduce the pollution by these backgrounds compared to Ref.~\cite{LHCb:2021trn}, resulting in a shift of around $0.064$. 
In addition, the explicit modeling of the residual backgrounds from misidentification in the fit results in a further shift of around $0.038$. 
The allowed statistical scatter for \RK\ is only $0.033$, for \RKst\ it is larger due to the significant increase of the data set. 

\section{Conclusions}
These proceedings report the most recent lepton flavour universality test in rare decays by the LHCb collaboration~\cite{LHCb:2022qnv,LHCb:2022zom}. 
The results are in excellent agreement with lepton flavour universality and supersede previous measurements~\cite{LHCb:2021trn,LHCb:2017avl}. 
The measurements are statistically dominated and 
will further improve in sensitivity with the data samples that LHCb will take during the Run~3 and beyond. 
While tensions with SM predictions in lepton universality tests of $\decay{b}{s\ellell}$ decays have disappeared, 
it should be noted that the anomalies in $\decay{b}{s\mumu}$ transitions are not affected by this change. 
However, as the SM predictions for these processes exhibit significant hadronic uncertainties,
more work is required on both the theory and experimental side to make progress on this question. 

\enlargethispage{1cm}
\section*{Acknowledgments}
C.\,L.\ gratefully acknowledges support by the 
Deutsche Forschungsgemeinschaft (DFG), 
grant identifiers LA 3937/1-1/2 and LA 3937/2-1.

\end{document}